\documentclass[letterpaper, preprint, paper,11pt]{AAS}	

\usepackage{bm}
\usepackage{amsmath}
\usepackage{siunitx}
\usepackage{subcaption}
\usepackage[colorlinks=true, pdfstartview=FitV, linkcolor=black, citecolor= black, urlcolor= black]{hyperref}
\usepackage{overcite}
\usepackage{graphicx}
\usepackage{footnpag}			      	

\PaperNumber{21-417}

\begin{document}

\title{Structural Design and Impact Analysis of a 1.5U CubeSat on the Lunar Surface}

\author{Christopher W. Hays\thanks{PhD Student, Aerospace Engineering Department, Embry-Riddle Aeronautical University, 1 Aerospace Blvd, Daytona Beach, FL., 32114},  
Daniel Posada\thanks{PhD Student, Aerospace Engineering Department, Embry-Riddle Aeronautical University, 1 Aerospace Blvd, Daytona Beach, FL., 32114},
Aryslan Malik\thanks{PhD Student, Aerospace Engineering Department, Embry-Riddle Aeronautical University, 1 Aerospace Blvd, Daytona Beach, FL., 32114},
Dalton Korczyk\thanks{Masters Student, Aerospace Engineering Department, Embry-Riddle Aeronautical University, 1 Aerospace Blvd, Daytona Beach, FL., 32114},
Ben Dafoe\thanks{Undergraduate Student, Aerospace Engineering Department, Embry-Riddle Aeronautical University, 1 Aerospace Blvd, Daytona Beach, FL., 32114},
and Troy Henderson\thanks{Associate Professor, Aerospace Engineering Department, Embry-Riddle Aeronautical University, 1 Aerospace Blvd, Daytona Beach, FL., 32114}
}

\maketitle{} 		

\begin{abstract}
Ahead of the United States' crewed return to the moon in 2024, Intuitive Machines, under a NASA Commercial Lunar Payload Services contract, will land their Nova-C lunar lander in October 2021. At 30 meters altitude during the terminal descent, EagleCam will be deployed, and will capture and transmit the first-ever third-person images of a spacecraft making an extraterrestrial landing. This paper will focus on the structural design, modeling, and impact analysis of a 1.5U CubeSat payload to withstand a ballistic, soft-touch landing on the lunar surface.

\end{abstract}


\section{Introduction}\label{sec: Introduction}
NASA's Commercial Lunar Payload Services (CLPS) contracts, part of the Artemis program, allow for multiple, rapid, commercial payload deliveries of instruments and technology demonstrations to the lunar surface. The first CLPS contracts were awarded to Intuitive Machines, Astrobotic, and Orbit Beyond\cite{Daines_2019}. Using these awards, both Intuitive Machines and Astrobotic have plans to carry a variety of payloads to lunar orbit and the lunar surface, providing valuable scientific research opportunities for lunar operations.

Intuitive Machines challenged Embry-Riddle Aeronautical University (ERAU) students to design, test, build, and operate a deployable camera to image the landing of their Nova-C spacecraft. EagleCam is the payload being developed, which will provide a full 360$^{\circ}$ field-of-view (FOV) of the spacecraft landing and landing site.
EagleCam will encompass a complete set of complimentary avionics including an inertial measurement unit (IMU) consisting of a 3-axis accelerometer and gyro, and five wide FOV cameras to provide valuable scientific data about the lunar surface environment. 

This paper focuses on the structural design and impact analysis of the 1.5U CubeSat payload after deployment and uncontrolled free-fall to the lunar surface, which the structure of the CubeSat must survive. The internal components must survive direct impact with the surface from an approximately 30 meter free-fall to successfully complete the mission. The CubeSat structural design is sketched and iterated using the CAD tool CATIA, with structural evaluation in LS-Dyna under a calculated load from a dynamic model of the deployment\cite{brown2002elements,hibbeler2015structural}. A preliminary structure was then manufactured and dropped into a Lunar Surface Environment Test Bed (LuSE).

To measure data from the impact, two different sets of data are taken: an IMU, providing acceleration measurements from which impact forces can be derived, and a high-speed camera, measuring the time and speed of impact, along with the deceleration from impact.
First, this paper will discuss an assumed dynamic scenario to drive the impact analysis and subsequent trade studies. Next, there will be a brief discussion on the structural design of the payload and Lunar Surface Environment Test Bed. Lastly, an impact analysis is performed in LS-Dyna with verified results from a drop-test of the payload.  

\section{Dynamics}\label{sec: Dynamics}

\begin{figure}[hbt!]
    \centering
    \includegraphics[width=\textwidth]{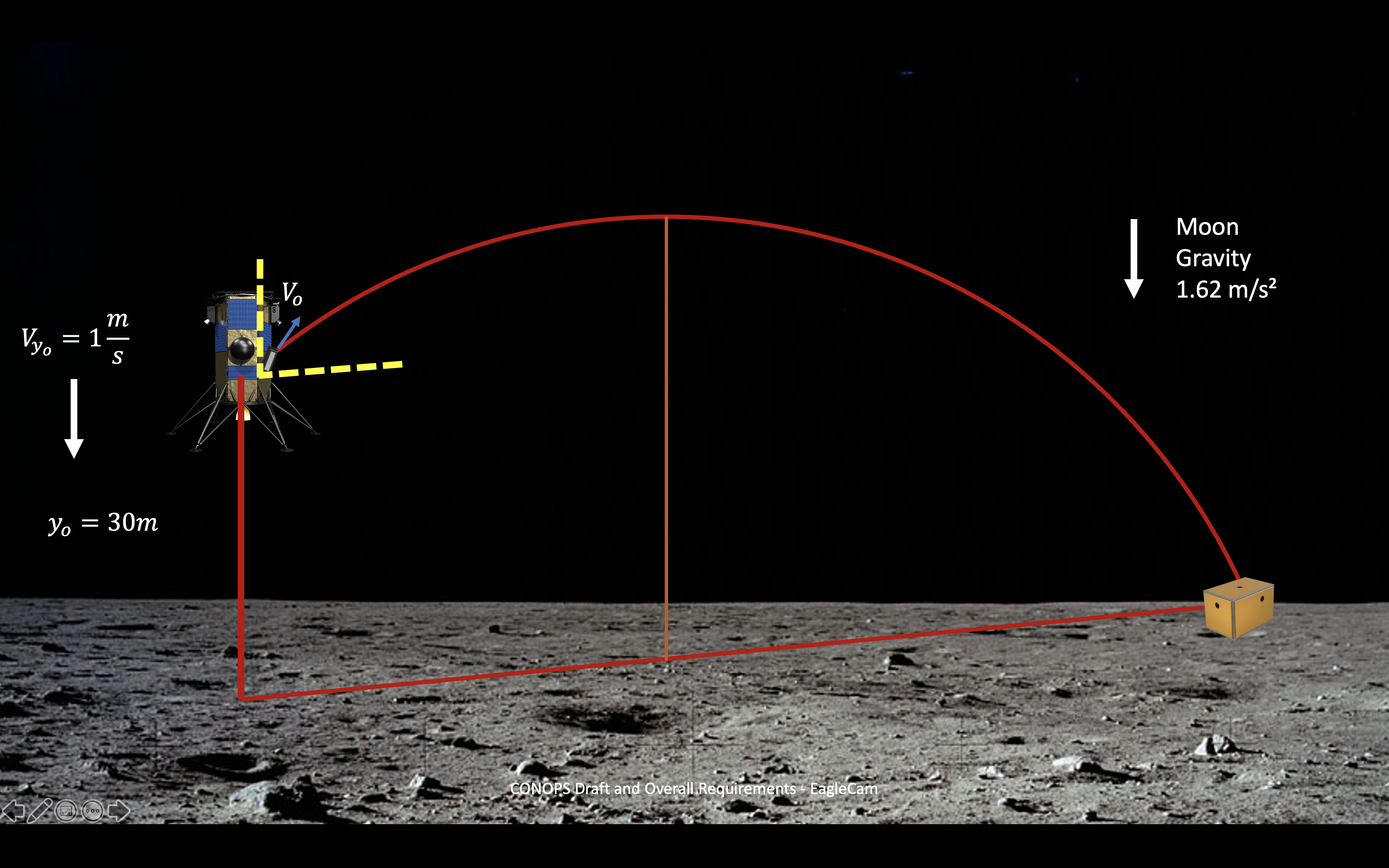}
    \caption{Nominal deployment during terminal phase.}
    \label{fig:dynamics}
\end{figure}

To make a realistic scenario, it is assumed that the payload is ejected in the terminal phase, or landing phase, from $30\,m$ above the lunar surface with an initial lander velocity of $1 \,m/s$. The impulse ejection provides the payload with an initial velocity of $\vec{v}_0=[0.5, ~ 2.1]^T ~m/s$ to reach a target landing range of $\approx$ 12.4 $m$. 
(The $12.4 m$ range was chosen to be clear of the lander's vehicle dispersion ellipse, but to also provide a good resolution image.) A depiction of this scenario can be seen in Figure \ref{fig:dynamics}.
Assuming lunar gravity to be $g=1.62$ $m/s^2$, the impact speed of the lander is to be $10.0926$ $m/s$.

Apart from kinematics and impact velocity, the orientation of the CubeSat when it reaches the lunar surface is of importance. The translational motion and rotational motion of EagleCam are assumed to be decoupled. Moreover, since there is no external torque applied (e.g., there is no atmosphere on the moon to perturb the motion), Euler's rotational equations\cite{schaub} may be applied in a straightforward fashion:
\begin{equation}\label{eulers}
\bm{I\dot{\omega}+\omega\times(I\omega)=0}
\end{equation}
For convenience, the inertia matrix of the CubeSat is obtained with respect to principal axes system from CATIA. Therefore, the equations simplify to the following:
\begin{align}\label{simpeuler}
\begin{split}
      I_{11}\dot{\omega}_{1}=-(I_{33}-I_{22})\omega_{2}\omega_{3}\\
      I_{22}\dot{\omega}_{2}=-(I_{11}-I_{33})\omega_{3}\omega_{1}\\
      I_{33}\dot{\omega}_{3}=-(I_{22}-I_{11})\omega_{1}\omega_{2}
\end{split}
\end{align}
Now, the ejector frame is defined to be aligned with the EagleCam's faces with $x$-axis pointing normal to the front face. Thus, rotation matrix from inertial frame to ejector frame is defined to be the following:
\begin{equation}\label{rot2e}
     R_{se}=\begin{bmatrix}
           \cos{\theta_{e}}&0&-\sin{\theta_{e}}\\
           0&1&0\\
           \sin{\theta_{e}}&0&\cos{\theta_{e}}\\
           \end{bmatrix}\in{SO(3)}
\end{equation}
where, $\theta_{e}$ is the ejection angle with respect to horizontal:
\begin{equation}\label{rot2e}
     \theta_{e}=\arctan{({v}_{0x}/{v}_{0y})}
\end{equation}
The principal axes system was obtained in CATIA. It was discovered that $x$-axes of ejector frame and body frame are the same. The rotation matrix from ejector frame to principal axes system ``body'' frame is given as:
\begin{equation}\label{rot2b}
     R_{eb}=\begin{bmatrix}
     1&0&0\\
           0&\cos{\phi}&-\sin{\phi}\\
           0&\sin{\phi}&\cos{\phi}\\
           \end{bmatrix}\in{SO(3)}
\end{equation}
where, $\phi$ is the angle about $x$-axis that rotates the axes system from ejector frame to body frame.
Given Equations \ref{rot2e} and \ref{rot2b}, the rotation matrix from inertial to body frame is obtained as follows:
\begin{equation}\label{s2b}
     R_{sb}=R_{se}R_{eb}\in{SO(3)}
\end{equation}
The preliminary inertia matrix given in the body frame is as follows:
\begin{equation}\label{inertia}
     \bm{I}=\begin{bmatrix}
           0.002&0&0\\
           0&0.003&0\\
           0&0&0.003\\
           \end{bmatrix}kgm^2
\end{equation}
It is assumed that there is an initial disturbance that gives initial condition to the angular velocity vector $\bm{\omega}(t=0)$. This vector is represented in the ejector frame, thus it should be transformed to the body frame so that Equation \ref{simpeuler} can be used to obtain the time history of $\bm{\omega}(t)$ vector.
\begin{equation}\label{omegab}
     \bm{\omega}_b=R_{eb}^T\bm{\omega}_e=R_{be}\bm{\omega}_e
\end{equation}
It should be noted that rotation matrices $R_{eb}$ and $R_{be}$ are constant.
Given time history of angular velocity vector, attitude kinematic equation is given as:
\begin{equation}\label{attitudekinematic}
 \dot{R}_{sb}=[\bm{\omega}]R_{sb}
\end{equation}
where,
\begin{equation}\label{skew}
 [\bm{\omega}]=\begin{bmatrix}
           0&-\omega_3&\omega_2\\
           \omega_3&0&-\omega_1\\
           -\omega_2&\omega_1&0\\
           \end{bmatrix}\in{so(3)}
\end{equation}
By integrating Equation \ref{attitudekinematic} the time history of body frame $R_{sb}(t)$ is obtained. The orientation of the EagleCam right before it lands on the lunar surface $R_{sb}(t_f)$ is used as an input to the FEA drop test simulation.
\begin{figure}[hbt!]
    \centering
    \includegraphics[width=\textwidth]{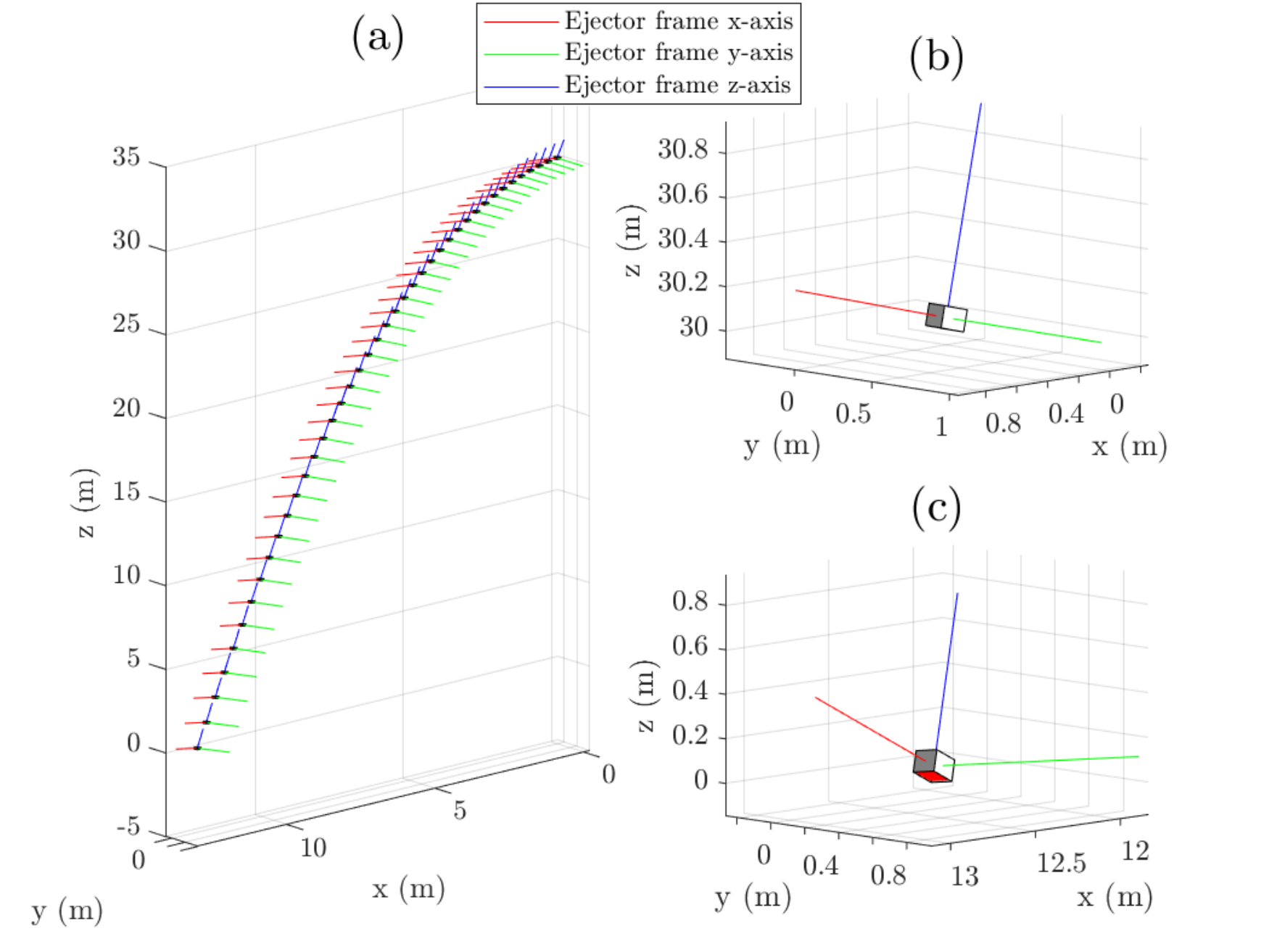}
    \caption{Dynamics simulation with $0.035$ $rad\,s^{-1}$ angular disturbance at the ejection with the full trajectory (a), first (b) and last (c) frames.}
    \label{fig:dynamics_sim}
\end{figure}

The dynamics of the ejection was simulated and the results are shown in Figure \ref{fig:dynamics_sim}, where it was assumed that there is a $0.035$ $rad\,s^{-1}$ disturbance at the ejection. It can be seen from the figure that the EagleCam lands on one of its vertices given the angular disturbance described above. The orientation of the EagleCam shown in Figure \ref{fig:dynamics_sim} (c) is used as an input to an LS-Dyna impact analysis simulation.
\section{Drop Module Structural Design}\label{sec: Drop Module}
The physical dimensions of the model are limited by the CubeSat launcher and necessary modification required for communications and power. The deployment device is selected to be a 3U launcher, but the intentions are to use some of the remaining space for the communications systems between the lander and the CubeSat. With these considerations, the maximum volume for the model is a 1.5U CubeSat. 

Taking inspiration from previous, successful CubeSats, the goal was to make a rigid frame with thin panels. This design is preferred for the purpose of mass savings, and also for manufacturability and ease of payload hardware integration. 

The preliminary design, to be used for preliminary structural modeling, is a simple square aluminum tubing with removable endcaps, machined from solid aluminum blocks. The solid aluminum frame means that the stress is able to move through the structure easier without interference from fasteners. The concept was to ensure proper modeling by matching the preliminary design impact model with the experiment, thereby gaining confidence in the simulated analysis of the flight model. The frame length is $150\,mm$, width and height are $100 \,mm$, and the thickness is $5/32\,in$. The design of the flight unit will contain mass-saving cut outs and Ultem panels to aid in the communication to the lander.

    \begin{figure}[h!]
        \centering
        \begin{minipage}[t]{.5\textwidth}
            \centering
            \includegraphics[width=\textwidth]{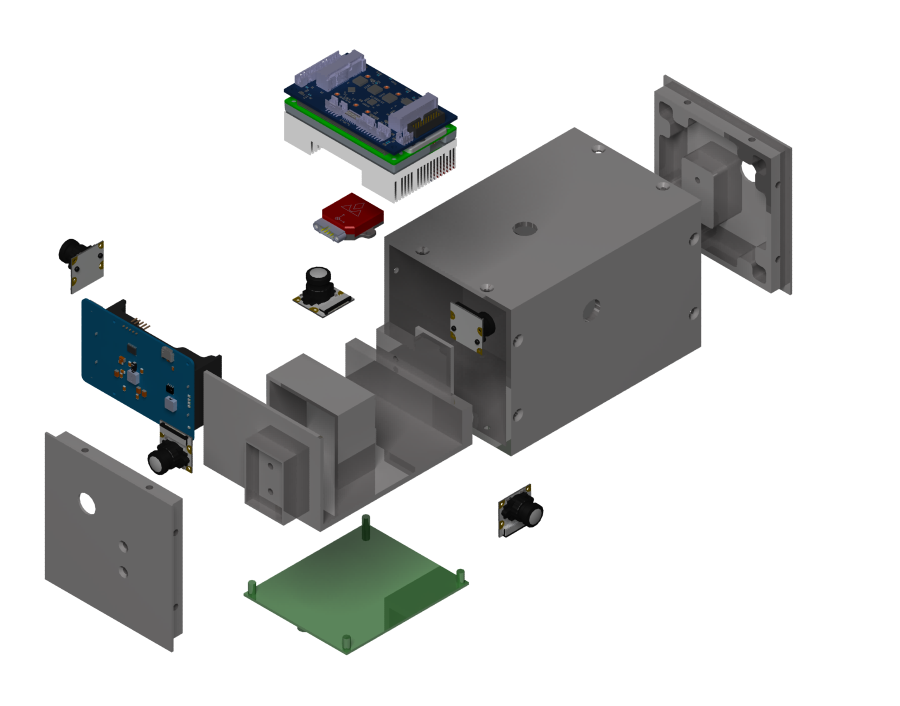}
            \subcaption{Exploded view of the drop test model structure.}
            \label{fig:exploded_view}
        \end{minipage}%
        \hfill
        \begin{minipage}[t]{0.5\textwidth}
            \centering
            \includegraphics[width=\textwidth]{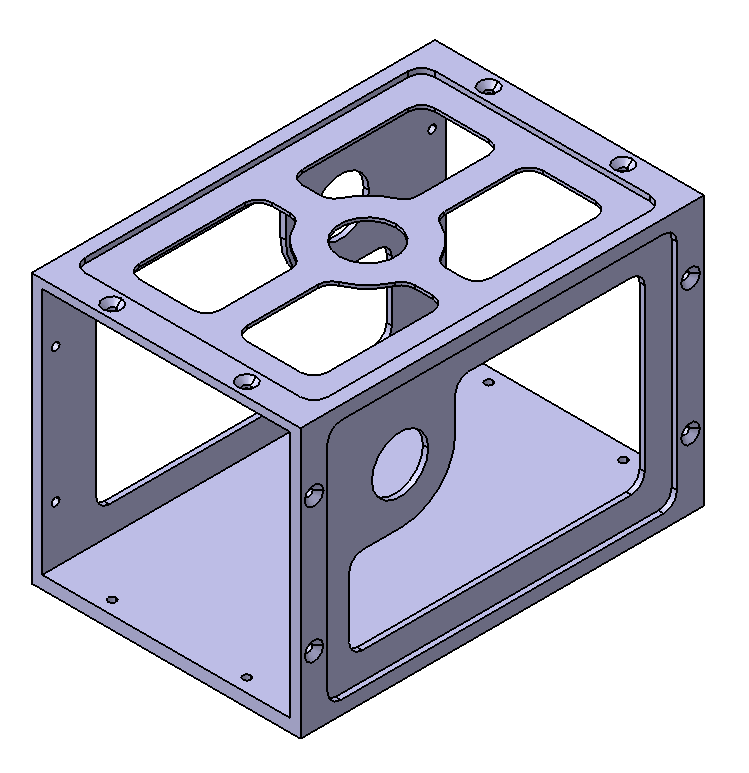}
            \subcaption{Revised external structure model for future use in the drop test.}
            \label{fig:drop_mod_structure}
        \end{minipage}
        \caption{Payload Drop Module External Structure.}
    \end{figure}
    
As seen in Figure \ref{fig:exploded_view}, all the components for the model are displayed in an exploded view. The model can fit five cameras, the Jetson TX2 with Elroy and heatsink (no fan), and the IMU.
The internal structure fits onto each endcap when assembled for minimal movement inside the shell with or without bolts to hold it in place. 
Unlike the main shell and endcaps, the internal structure is made from PLA with a 3D printer. Instead of estimating where the mounting holes may line up, they are placed over an additional PLA panel. 
Once in hand, the holes are drilled for fitting. The placement of components on the internal structure is mainly done inside the CAD software CATIA where there are models of each component. 
The main goal is to fit each component, the second to keep the IMU in the center of the body, and if possible to make it balanced along its main axis. 
In order to maximize the field of view, the lenses of the cameras should be protruding from the structure slightly to ensure the full 180$^{\circ}$ view. 
To accomplish this with a slide-in structure, the camera is mounted on the internal structure, slid in, and then each lens is attached through the shell.


The model used in the first physical test included the following elements: NVIDIA Jetson TX2 on board computer, Connect Tech Elroy carrier board, VectorNav VN-100 inertial measurement unit, Geekworm x728 power supply unit, an in-house voltage upscale PCB, and two Panasonic 18650 batteries.
\section{Lunar Surface Environment Test Bed}\label{sec: LuSE}

The Lunar Surface Environment Test Bed (LuSE) was constructed to simulate the lunar landing, to capture the impact of the payload, and to image resulting regolith dispersion. 
LuSE's structure was divided into two primary components: the drop-module capture bay, and the sensor bench.

\subsection{Capture Bay}\label{sec: Capture Bay}

The LuSE Capture Bay is a 0.71 $\times$ 0.96 $\times$ 0.79 $m$ confinement enclosed with four plexiglass walls with additional L-bracket reinforcement. 
Two of these walls are set up for viewing for the sensor bench, detailed in the next section, one wall contains an image calibration checkerboard, and the fourth wall is painted black to reduce the amount of incoming light and glare.

\begin{figure}
    \centering
    \includegraphics[width=0.55\textwidth]{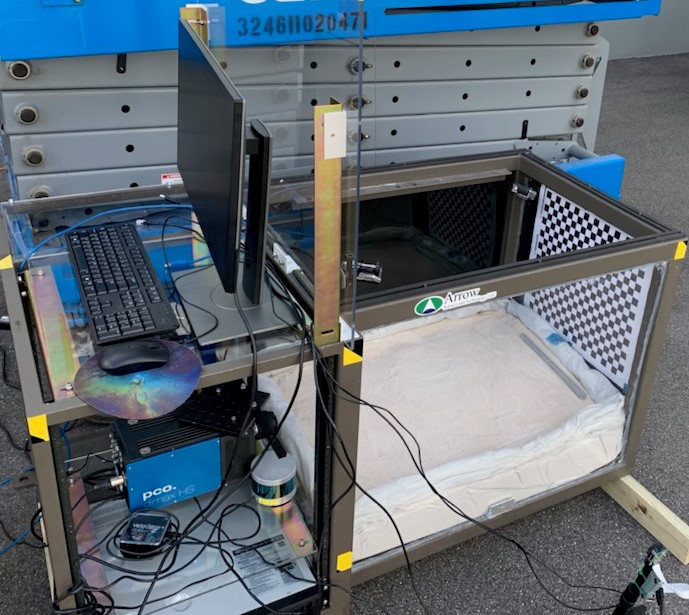}
    \caption{Lunar Surface Environment Test Bed Capture Bay and Sensor Bench.}
    \label{fig:capture_bay}
\end{figure}

The bottom 0.15 $m$ of the bay is filled with a combination of rocks and sand to simulate the lunar surface. \cite{compston1970chemistry}
Future drop tests will include the use of lunar regolith simulant, manufactured by the CLASS Exolith Lab at the University of Central Florida, \footnote{CLASS Exolith Lab -  https://sciences.ucf.edu/class/exolithlab/} to provide a more accurate representation of the surface impact at landing.
Figure \ref{fig:capture_bay} shows the full LuSE setup during the drop test.

\subsection{Sensor Bench}\label{sec: Sensor Bench}

The Sensor Bench on-board LuSE is comprised of multiple GoPro cameras, two FLIR Blackfly S machine vision cameras, one Raspberry Pi High Quality camera, a Velodyne Puck LITE Lidar, and a pco.dimax HS4 high-speed camera. 
Each sensor was include to capture a unique perspective of the drop test and maximize scientific return from each drop performed. 
Each of the GoPros were setup to capture the drop test from the top-down from different angles. 
One GoPro was attached to the top of LuSE on the plexiglass shield, and one was taken up on the scissor lift to provide a perspective from drop apogee. 
The Raspberry Pi High-Quality camera took a full 180$^{\circ}$ fov of the drop test. 
The lidar was included to analyze lidar performance in a dust cloud and to evaluate what valuable data could be drawn out.
To capture a slow-motion video of the drop to perform impact analysis, the pco.dimax HS4 high-speed camera was considered the primary sensor on the bench.
The two FLIR Blackfly S machine vision cameras were used as backups in case none of the other views captured good footage.

    \begin{figure}[h!]
        \centering
        \begin{minipage}[t]{.43\textwidth}
            \centering
            \includegraphics[width=\textwidth]{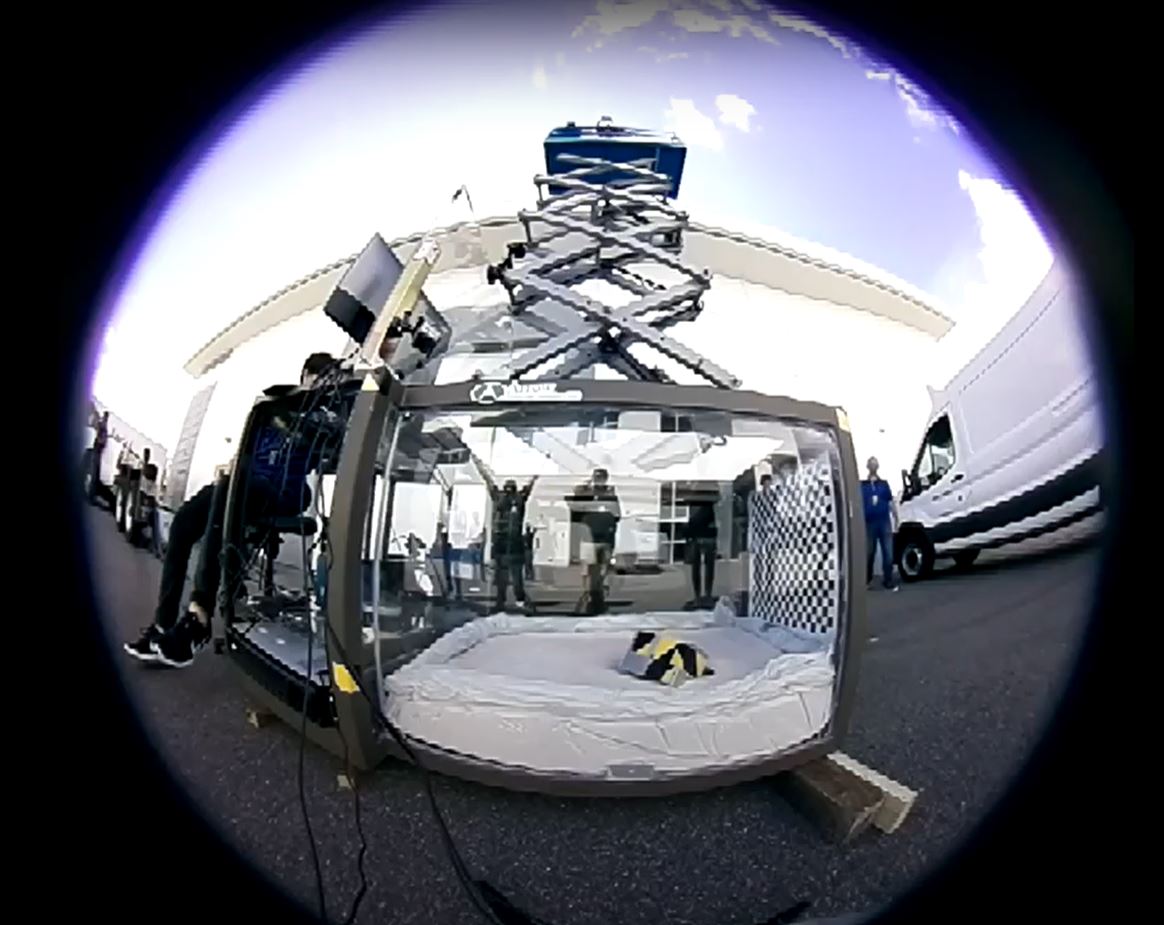}
            \subcaption{Raspberry Pi High Quality Camera.}
            \label{fig:RPi_camera}
        \end{minipage}%
        \hfill
        \begin{minipage}[t]{0.5\textwidth}
            \centering
            \includegraphics[width=\textwidth]{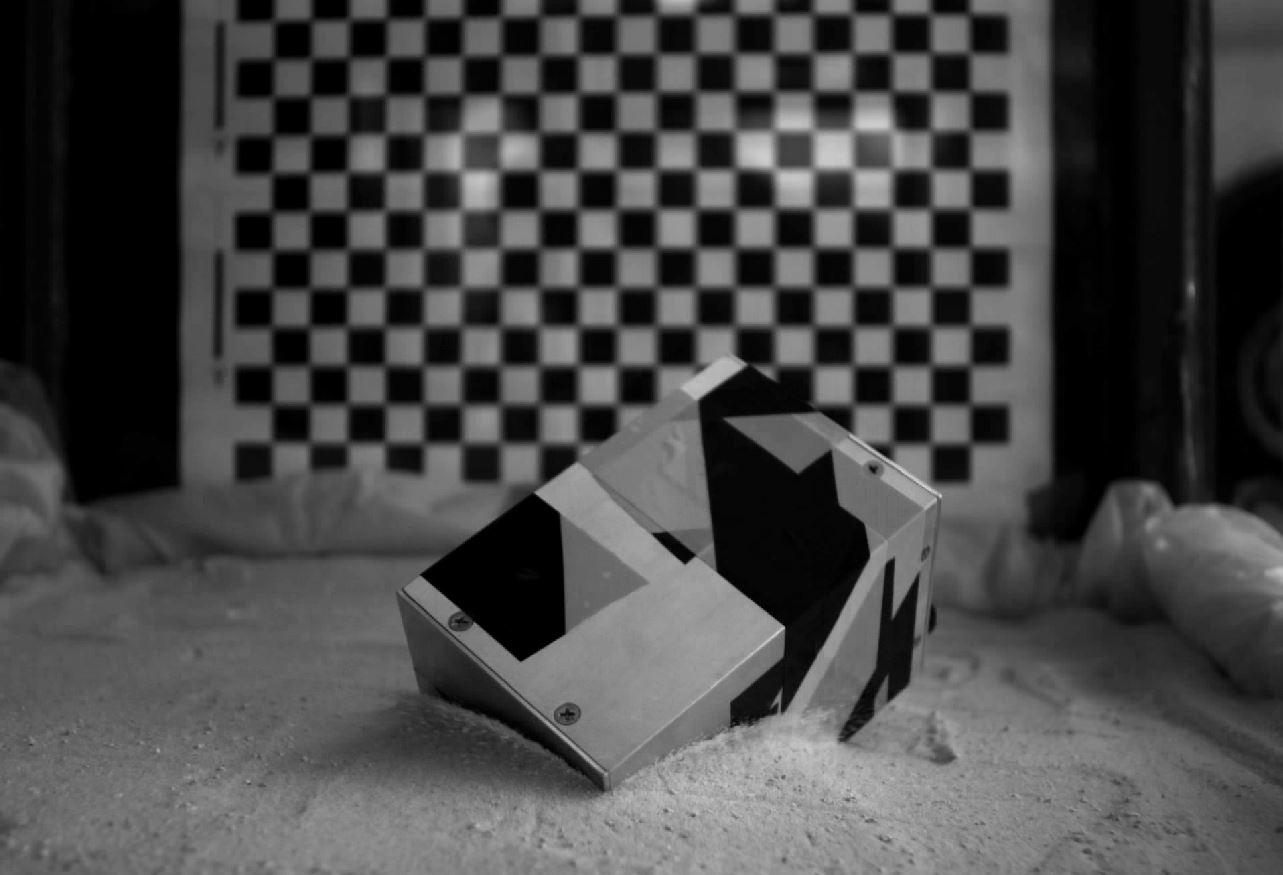}
            \subcaption{pco.dimax HS4}
            \label{fig:pco_image}
        \end{minipage}
        \caption{Sample imagery from select sensors on LuSE.}\label{fig:sample_imagery}
    \end{figure}

\section{Results}\label{sec: Results}

This section presents the Finite Element Analysis (FEA) setup used to perform the impact analysis. 
The simulation construction and parameters are discussed, as well as how the initial conditions are obtained. 
Next, experimental and simulated results are compared for simulation validation. 
Finally, results from a full-scale drop test are presented to verify impact velocity and results from the simulation of a full-scale test are extrapolated.

\subsection{FEA Setup}\label{sec: FEA Results}

In order to perform a simulation of surface impact, LS-Dyna's impact analysis capability was utilized. The simulations were set up for the purpose of model validation and all of the parameters were initialized to represent the experimental drop tests. 

\begin{table}[htb!]
\centering
\caption{Mat 5 input for the sand simulation}
\label{tab:dyna:mat5}
\resizebox{0.65\textwidth}{!}{%
\begin{tabular}{l|l|l|l}
\hline
\multicolumn{1}{c|}{\textbf{Variable}} & \multicolumn{1}{c|}{\textbf{LS-Dyna Name}} & \multicolumn{1}{c|}{\textbf{Value}} & \multicolumn{1}{c}{\textbf{Units}}\\ \hline
Density & RO & 1.5E-3 & $g/mm^3$  \\
Shear Modulus & G & 1.5 & $MPa$  \\
Bulk Unloading Modulus & BULK & 60 & $MPa$  \\
Yield Surface Coefficient & A0 & 0 & $MPa^2$  \\
Yield Surface Coefficient & A1 & 0 & $MPa$  \\
Yield Surface Coefficient & A2 & 0.3 & N/A  \\
Pressure Cutoff & PC & 0 & $MPa$  \\
Crushing Option & VCR & 0 (default) & N/A \\
Reference Geometry & REF & 0 (default) & N/A \\\hline
\end{tabular}%
}
\end{table}

The following assumptions were made: for the preliminary simulation only the shell of the CubeSat was simulated, the material of the shell was assumed to be Aluminum 6061-T6, the gravity was assumed to be $9.81$ $m/s^2$ and it is applied through \textbf{LOAD\_BODY\_Z} card, the shell's initial velocity was initialized to $6.0$ $m/s$ which was obtained from kinematics (the initial height of the drop test is approximately $1.83$ $m$) and its position was $80$ $mm$ above the sand, and it was oriented at $0^{\circ}$ with respect to horizontal, the sand was simulated as a soil/foam material. 

\begin{figure}[htb!]
    \centering
    \includegraphics[scale= 0.4]{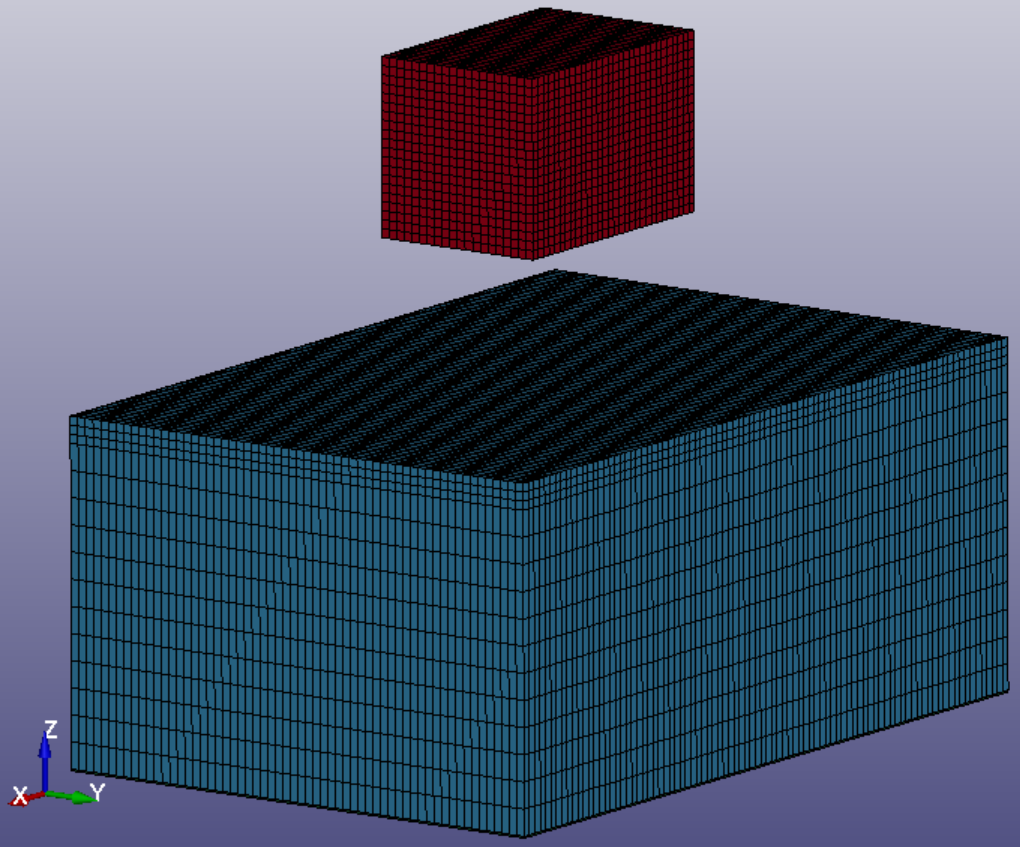}
    \caption{LS-Dyna simulation setup.}
    \label{fig:sim_setup}
\end{figure}

The simulation setup is shown in Figure \ref{fig:sim_setup}, where the EagleCam outer shell is shown in red, and the sand is shown in blue. The shell was modeled as elastic material with the following properties: density $\rho=0.00266$ $g/mm^3$, Young's modulus $E=6895$ $MPa$, and Poisson's ratio $\nu=0.325$. The dimensions of the sand are $450$ $mm\times{}300$ $mm\times195$ $mm$, and it was modeled as Mat 5, \textbf{SOIL$\_$AND$\_$FOAM}, with card values shown in Table \ref{tab:dyna:mat5} which were calibrated based on the experimental data provided by NASA technical report\cite{lsdynasoil,lsdynasoil2, lsdynamanual}. In addition, a node was placed at the position of the IMU and it was constrained using \textbf{CONSTRAINED\_INTERPOLATION\_LOCAL} card to all nodes of the EagleCam shell. A couple of nodes at the top of the shell were used to record the acceleration response as well using \textbf{DATABASE\_HISTORY\_NODE\_ID} card.

\begin{figure}[htb!]
    \centering
    \includegraphics[scale= 0.5]{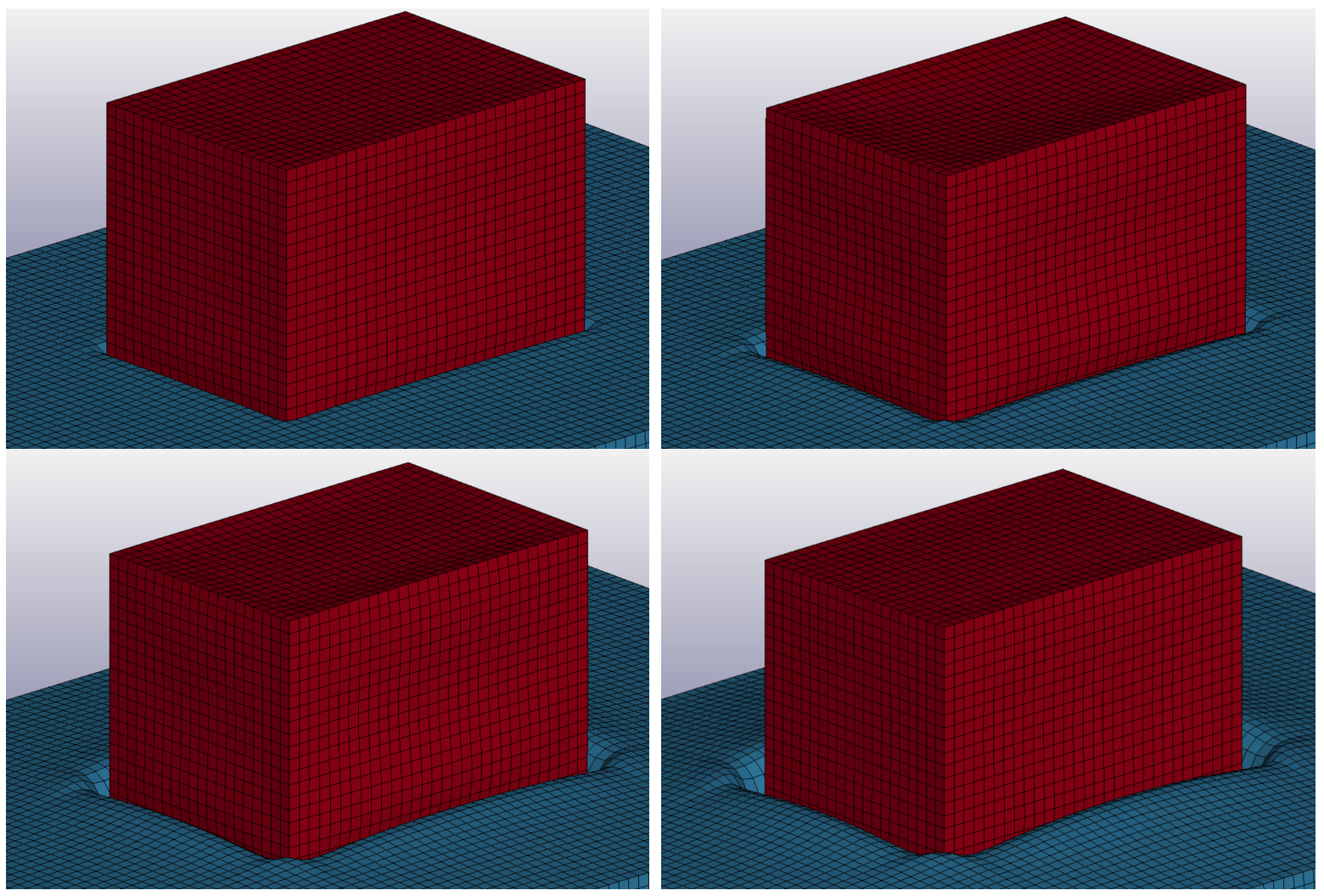}
    \caption{LS-Dyna simulation snapshots.}
    \label{fig:sim_snapshots}
\end{figure}

Moreover, hydrostatic pressure in the sand due to gravity acting on an overburden material was taken into account. The lower surface of the sand was restricted in all translational and rotational components using SPC Boundary card. Also, damping was introduced to the simulation through \textbf{DAMPING\_GLOBAL} card, and with the activation happening after the impact which happens at $14$ $ms$ in this simulation. 

\begin{figure}[h!]
        \centering
        \begin{minipage}[t]{.41\textwidth}
            \centering
            \includegraphics[width=\textwidth]{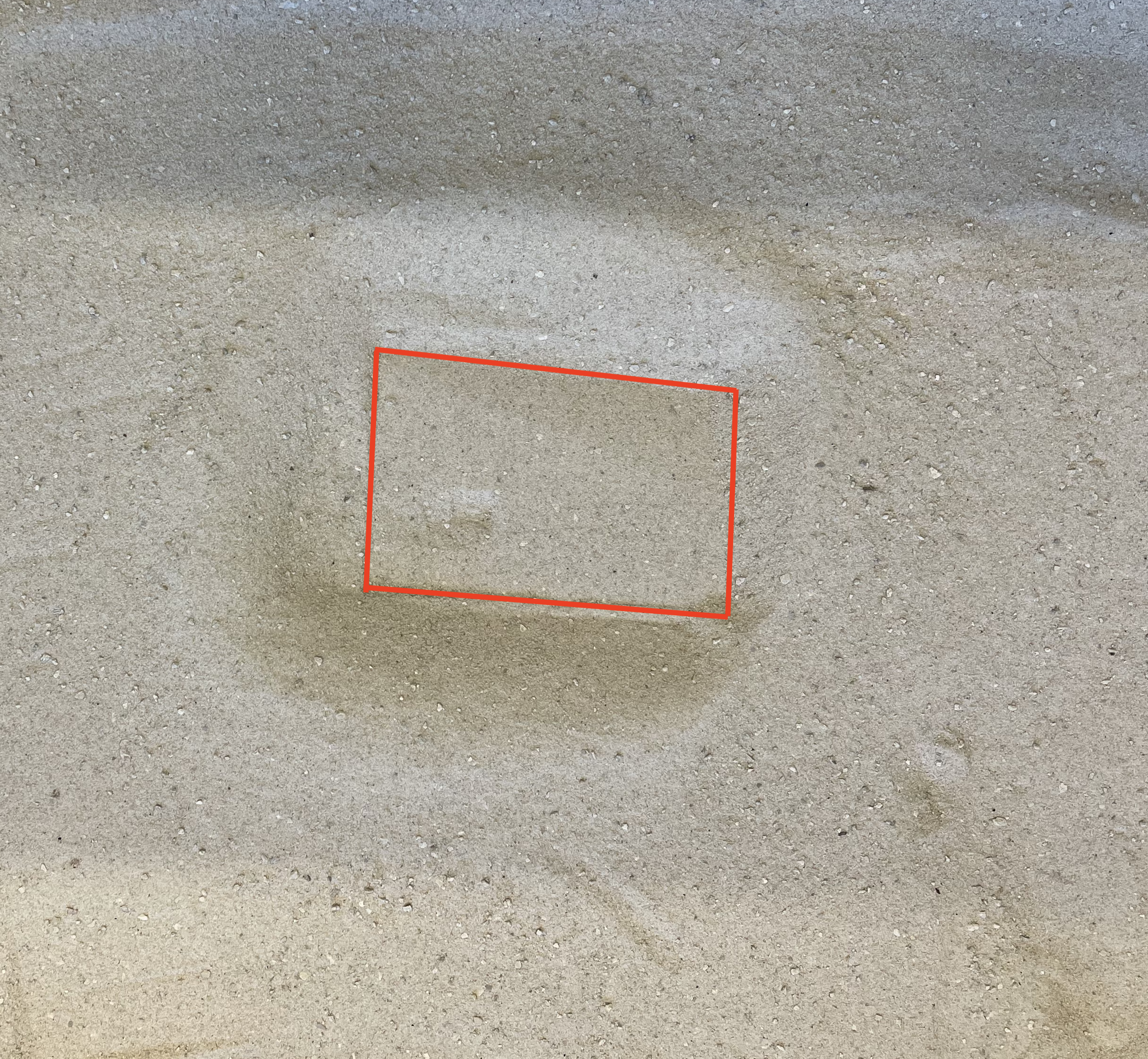}
            \subcaption{Top view.}
            \label{fig:crater_comparison:top}
        \end{minipage}%
        \hfill
        \begin{minipage}[t]{0.5\textwidth}
            \centering
            \includegraphics[width=\textwidth]{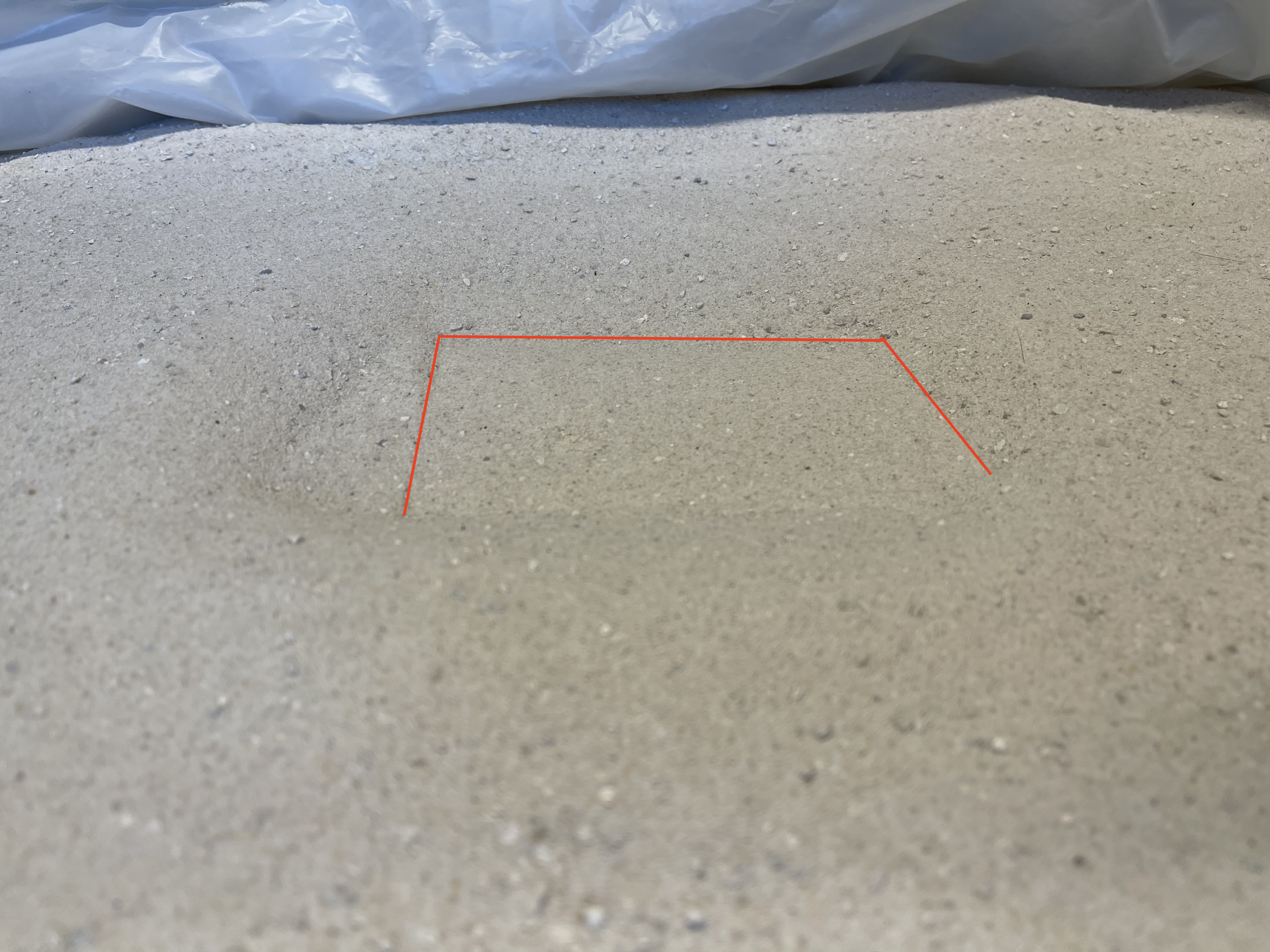}
            \subcaption{Side isometric view}
            \label{fig:crater_comparison:side}
        \end{minipage}
        \hfill
        \begin{minipage}[t]{.41\textwidth}
            \centering
            \includegraphics[width=\textwidth]{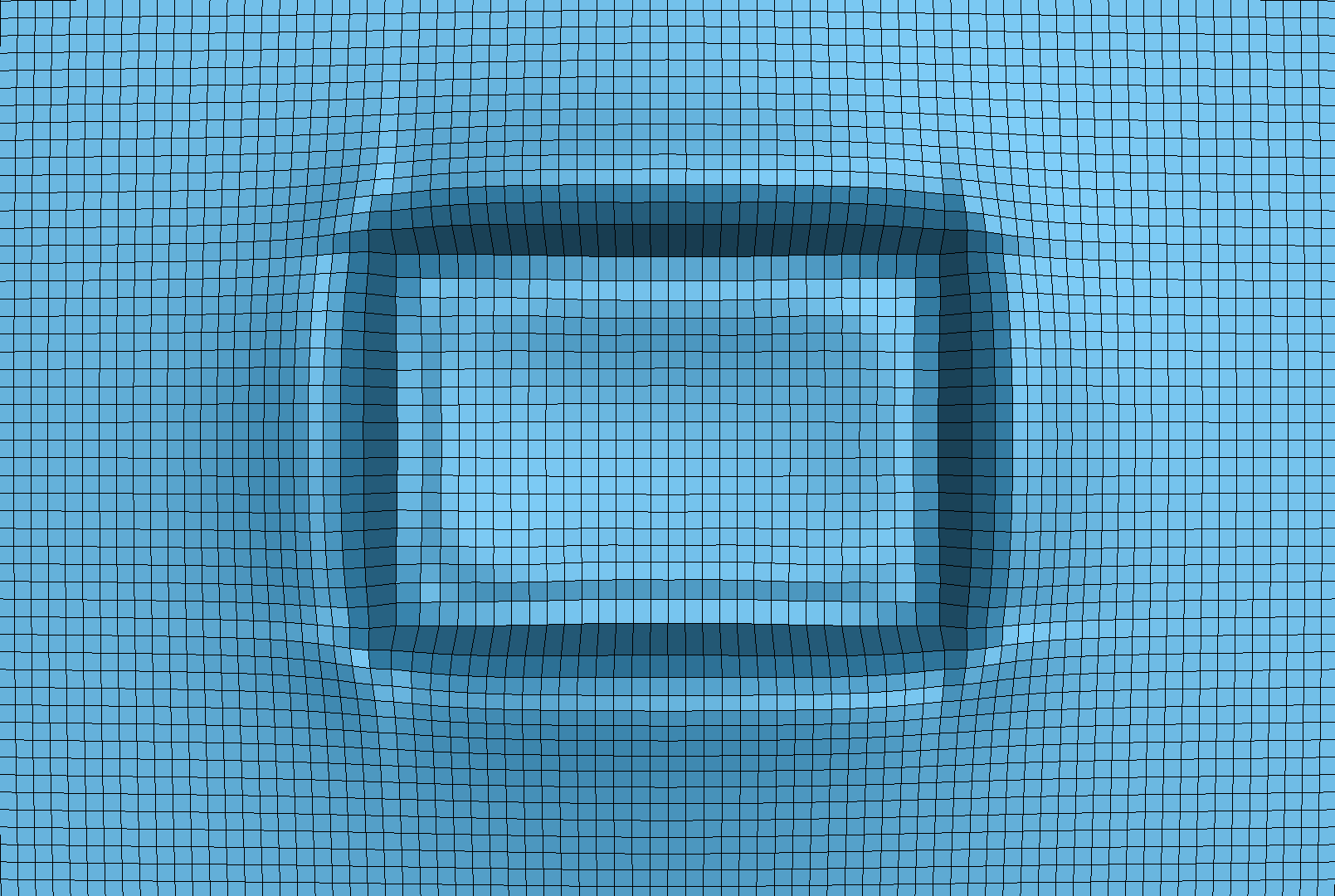}
            \subcaption{Sim top view.}
            \label{fig:crater_comparison:sim:top}
        \end{minipage}%
        \hfill
        \begin{minipage}[t]{0.5\textwidth}
            \centering
            \includegraphics[width=\textwidth]{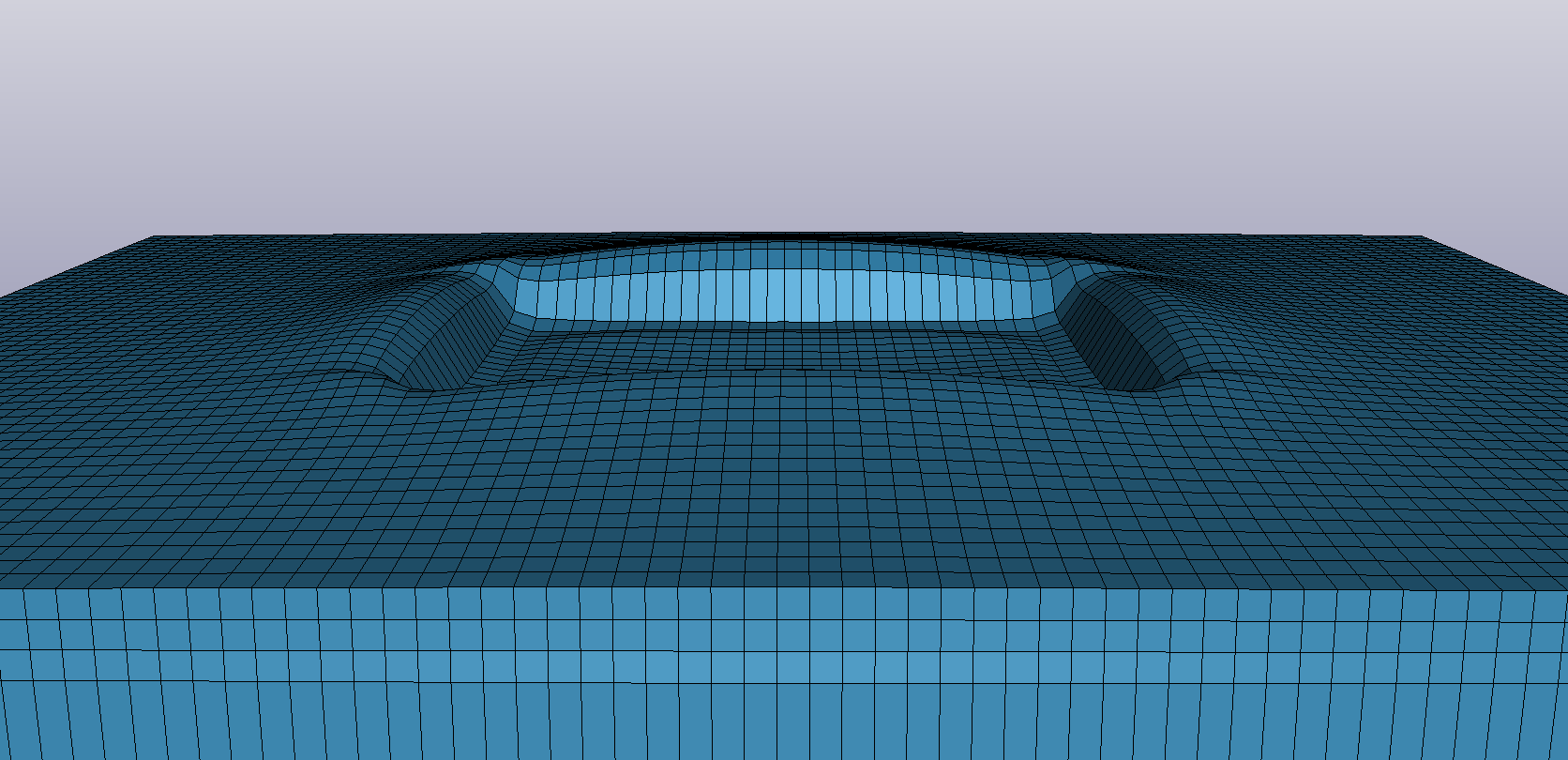}
            \subcaption{Sim side isometric view}
            \label{fig:crater_comparison:sim:side}
        \end{minipage}
        \caption{Impact crater comparison after surface impact on LuSE and LS-Dyna.}\label{fig:crater_comparison}
    \end{figure}

The snapshots of the simulation are shown in Figure \ref{fig:sim_snapshots}, where the forming of the impact crater can be seen. The maximum depth of the crater in simulation was approximately $17-18$ $mm$. The impact crater obtained in the simulation correctly resembles the impact crater from the experiments, this is demonstrated in Figure \ref{fig:crater_comparison}. Overall, the resemblance of the shape and depth of the impact crater validates the accuracy of the simulation model.

\begin{figure}[h!]
    \centering
    \includegraphics[width=0.6\textwidth]{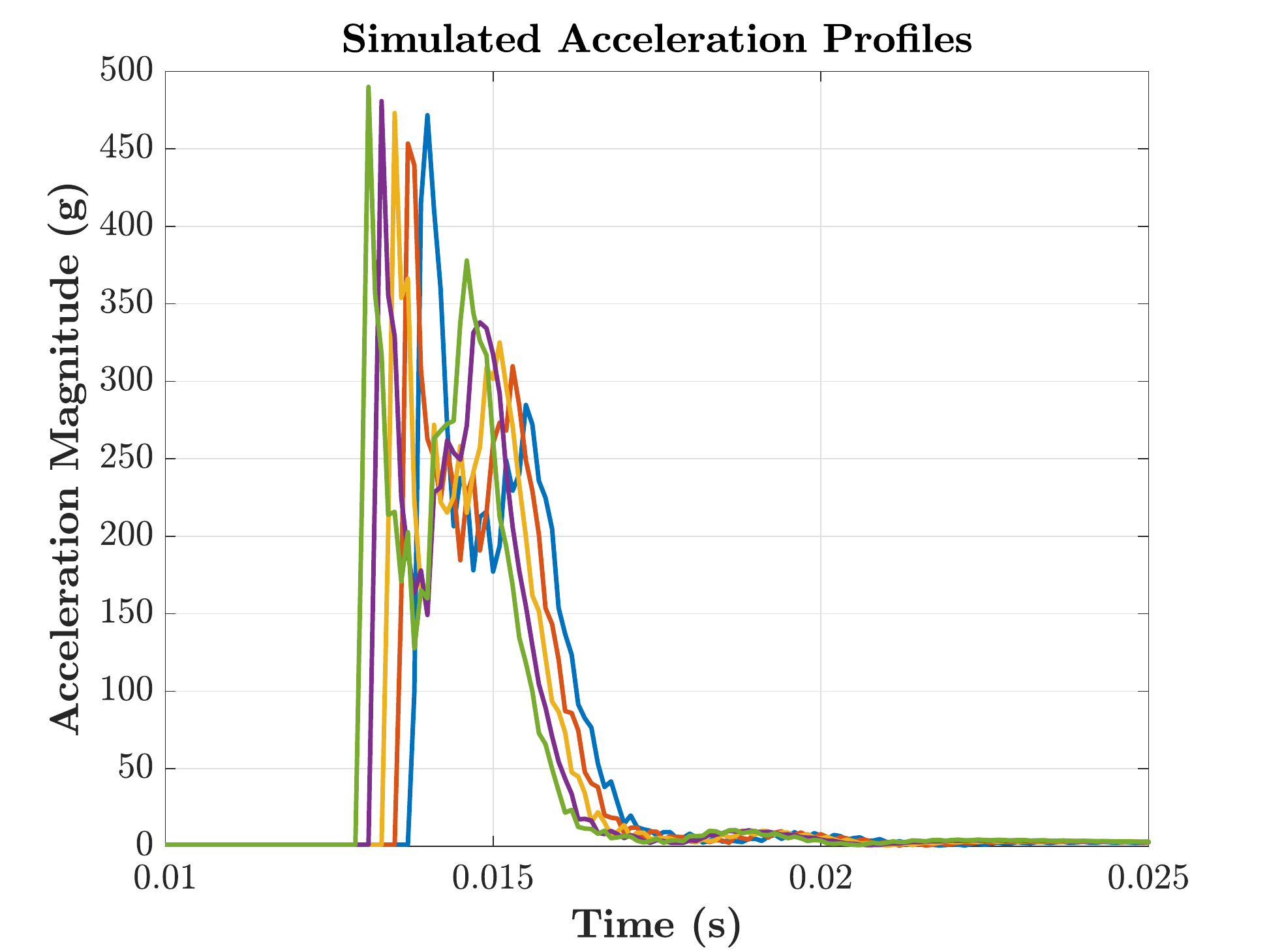}
    \caption{LS-Dyna Simulation acceleration profile results. Accelerations were taken at placement of IMU.}
    \label{fig: sim_accel_profiles}
\end{figure}

The velocity initial condition was varied from $6.0$ $m/s$ to $6.5$ $m/s$ with an increment of $0.1$ $m/s$ to see how it affects the acceleration response profile of the EagleCam shell.
Figure \ref{fig: sim_accel_profiles} shows the simulated acceleration results of EagleCam through impact. 
The LS-Dyna measurement node used to obtain the accelerations was placed at the IMU location in the final design. 
These results will be compared to the experimental results obtained from the drop tests. 

\subsection{Drop Test Results}\label{sec: Drop Test Results}

To evaluate the efficacy of the LS-Dyna simulations, a series of small-scale drop tests were performed.
Due to space limitations and feasibility issues associated with full-scale drop tests these small-scale tests were limited to a height of 7 feet from the top of the sand inside the capture bay of LuSE. 
It is assumed that if the simulation matches the experimental results, the simulation results may be extrapolated for full-scale tests. 

\begin{figure}[h!]
    \centering
    \includegraphics[width=0.6\textwidth]{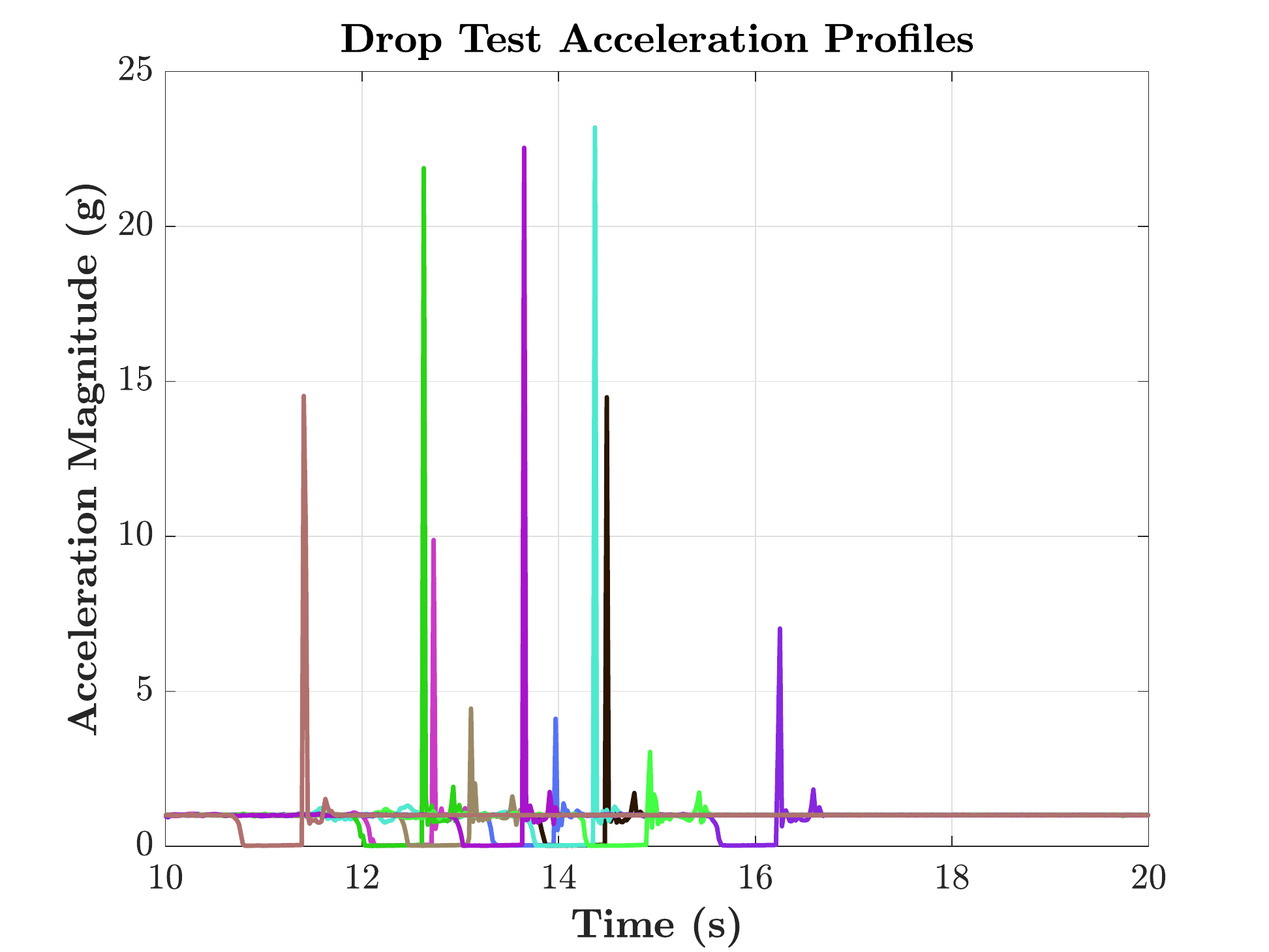}
    \caption{Acceleration profiles of EagleCam from IMU during small-scale drop tests.}
    \label{fig: exp_accel_profiles}
\end{figure}

To perform these small-scale drops, the drop module was equipped with an IMU placed in the anticipated final location on EagleCam. 
The drop module was then dropped from a height of 7 ft into the capture bay - resulting in an initial impact velocity of 6.469 $m/s$.
The results of each of these drop tests are shown in Figure \ref{fig: exp_accel_profiles}

\begin{figure}
    \centering
    \includegraphics[width=0.6\textwidth]{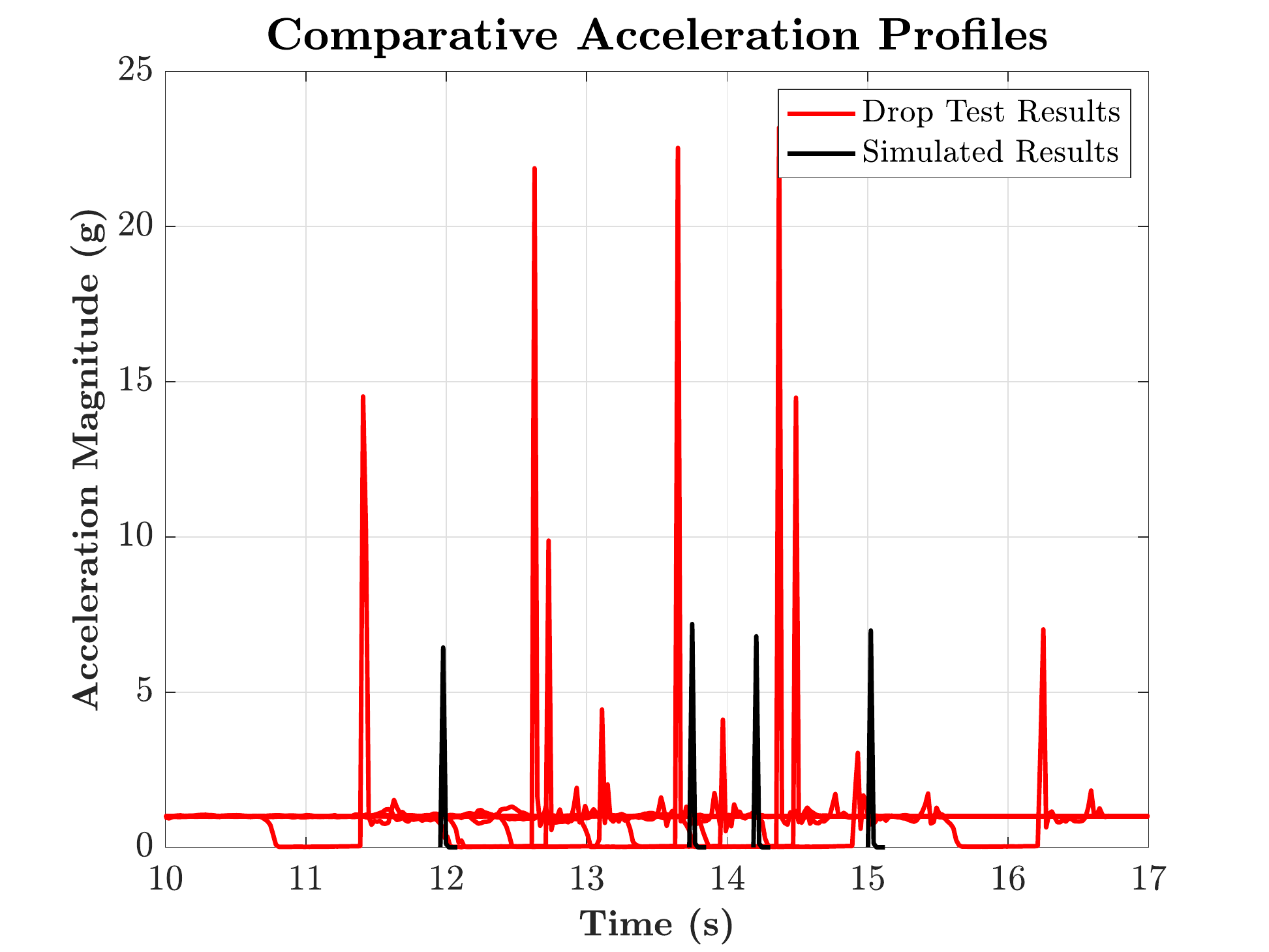}
    \caption{Comparison of the Simulated and Drop Test results.}
    \label{fig: comp_accel_profiles}
\end{figure}

At first glance, the acceleration profiles obtained in Figures \ref{fig: sim_accel_profiles} and \ref{fig: exp_accel_profiles} differ by close to an order of magnitude. 
This discrepancy is a result of the time step used for the simulation and the sampling frequency of the IMU.
The simulation used a time step of $0.1$ $ms$ which equates to $10$ $kHz$, while the IMU sampled at a frequency of $50$ $Hz$.
To appropriately validate the simulation, the simulated acceleration profiles were down-sampled to match the $50$ $Hz$ frequency of the IMU.
Figure \ref{fig: comp_accel_profiles} presents a comparison of the drop test acceleration profiles and the down-sampled simulated acceleration profiles. 

The drop test impact acceleration has a value $8.85 \pm 6.01$ $g$ with the simulated impact acceleration of $5.49 \pm 1.74$ $g$. 
Based on these results, it can be concluded that the true mean impact acceleration could lie within the simulated data.  
The discrepancies seen in the results could be contributed from a number of different sources. 
Primarily these sources are attributed to uncontrollable factors associated with the drop tests. 
Sources such as an inconsistent drop height hovering around 7 $ft$, an added initial velocity at drop, and an added angular velocity at drop each of which would contribute to an increased and widely varying impact velocity.
The elasticity of the EagleCam module also plays a role in shifting the results from the nominal, primarily contributing to the shifting of vibrational modes within the structure. 
Furthermore, by inspecting experimental data it was noticed that the acceleration in $x$ and $y$ axes contribute a lot to the total magnitude of the acceleration response, whereas in the simulation the main contributor is the acceleration in the $z$ axis and the rest are negligible. This can be explained by the fact that in the simulation the EagleCam shell falls perfectly flat on the bottom surface and in experiments there is some misalignment and the EagleCam falls at some arbitrary orientation due to initial disturbance. The lack of $x$ and $y$ acceleration components contribute to the difference in the acceleration response magnitude between the drop test experiments and the simulation as shown in Figure \ref{fig: comp_accel_profiles}.

\subsection{High-Speed Video Analysis of Large Drop}
\begin{figure}[hbt!]
    \centering
    \includegraphics[width=0.5\textwidth]{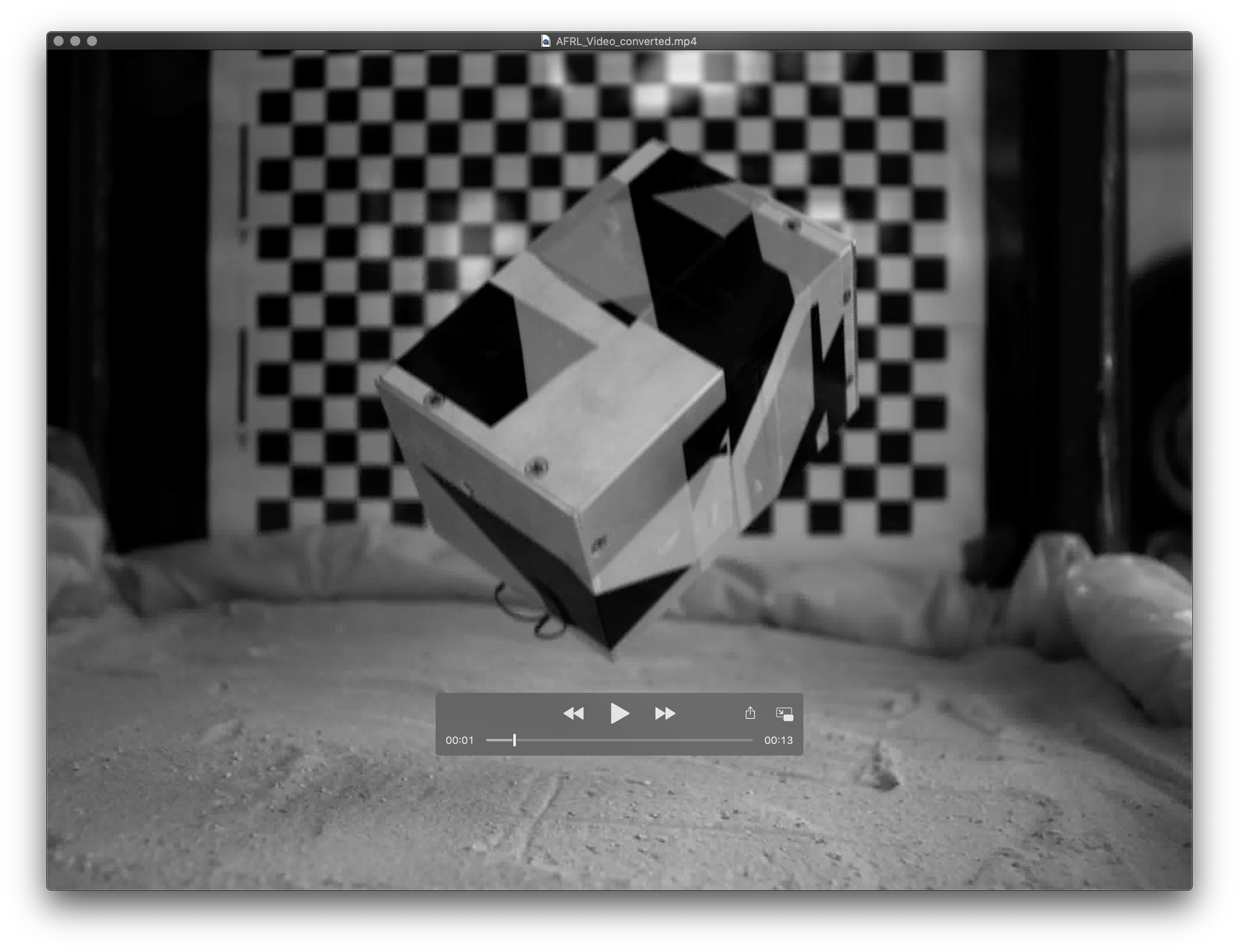}
    \caption{High-Speed Camera sample frame before impact}
    \label{fig:hispeed:frame}
\end{figure}

\begin{figure}[hbt!]
    \centering
    \includegraphics[width=0.8\textwidth]{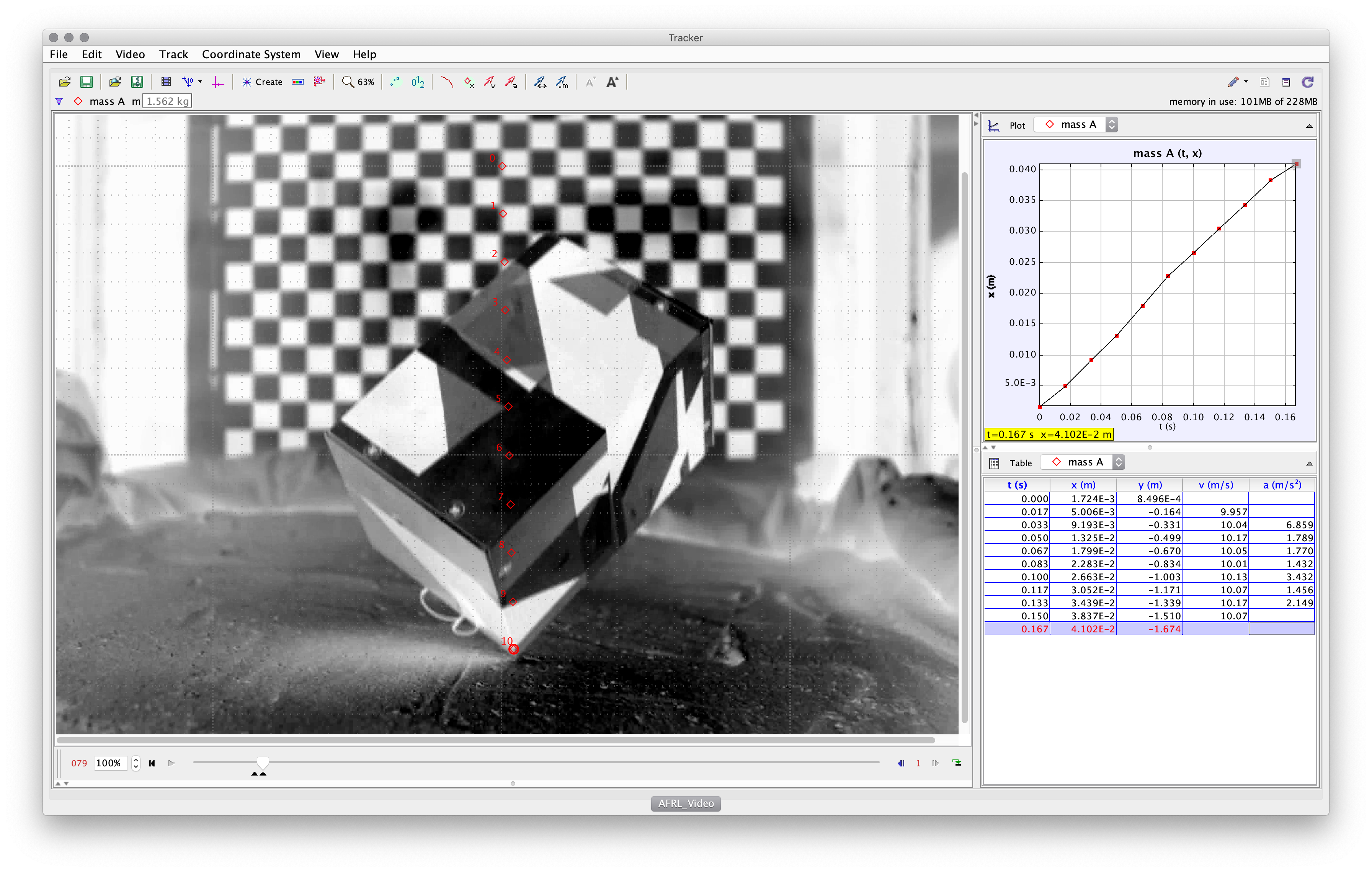}
    \caption{Feature tracking before impact to measure impact speed.}
    \label{fig:hispeed:tracker}
\end{figure}

To analyze the high-speed video from the drop test, the open source software Tracker was used to track a feature and get an approximate real position, speed, and acceleration with a calibration process.
Figure \ref{fig:hispeed:frame} shows one of the frames used to track the feature in the drop test module in order to obtain the speed. 
Figure \ref{fig:hispeed:tracker} shows the output from Tracker, where the feature tracked is seeing in read accompanied by numbers in the main screen. 
The output of the feature tracked provided the data in Table \ref{tab:tracker:output}.

\begin{table}[htb]
\centering
\caption{Tracker Output}
\label{tab:tracker:output}
\resizebox{0.5\textwidth}{!}{%
\begin{tabular}{l|l|l|l|l}
\hline
\multicolumn{1}{c|}{\textbf{time $s$}} & \multicolumn{1}{c|}{\textbf{x $m$}} & \multicolumn{1}{c|}{\textbf{y $m$}} & \multicolumn{1}{c|}{\textbf{v $m/s$}} & \multicolumn{1}{c}{\textbf{a $m/s^2$}} \\ \hline
0.000 & 1.724E-3 & 8.496E-4 & - & - \\
0.017 & 5.006E-3 & -0.164 & 9.957 & - \\
0.033 & 9.193E-3 & -0.331 & 10.04 & 6.859 \\
0.050 & 1.325E-2 & -0.499 & 10.17 & 1.789 \\
0.067 & 1.799E-2 & -0.670 & 10.05 & 1.770 \\
0.083 & 2.283E-2 & -0.834 & 10.01 & 1.432 \\
0.100 & 2.663E-2 & -1.003 & 10.13 & 3.432 \\
0.117 & 3.052E-2 & -1.171 & 10.07 & 1.456 \\
0.133 & 3.439E-2 & -1.339 & 10.17 & 2.149 \\
0.150 & 3.837E-2 & -1.510 & 10.07 & - \\ \hline
\end{tabular}%
}
\end{table}

Furthermore, from the video analysis in Figure \ref{fig:hispeed:tracker} it was obtained that the impact lasted around 10 frames at $0.017 s$ per frame. 
This time and the average impact speed provides an estimate of the impact force by obtaining the change in momentum. Recall Newton's second law:

\begin{equation}
    \Vec{F} = \frac{\Delta\Vec{p}}{\Delta t}
\end{equation}
where $\Delta\Vec{p}$ is the change of momentum, and $\Delta t$ the duration of this exchange.
\begin{equation}
    \Vec{F} = \frac{m\Delta \Vec{v}}{\Delta t}
\end{equation}
$$|\Vec{F}| = \left|\frac{m(v_2 - v_1)}{\Delta t}\right|$$
$$F \approx 92.66 N$$

\begin{figure}
    \centering
    \includegraphics[width=0.6\textwidth]{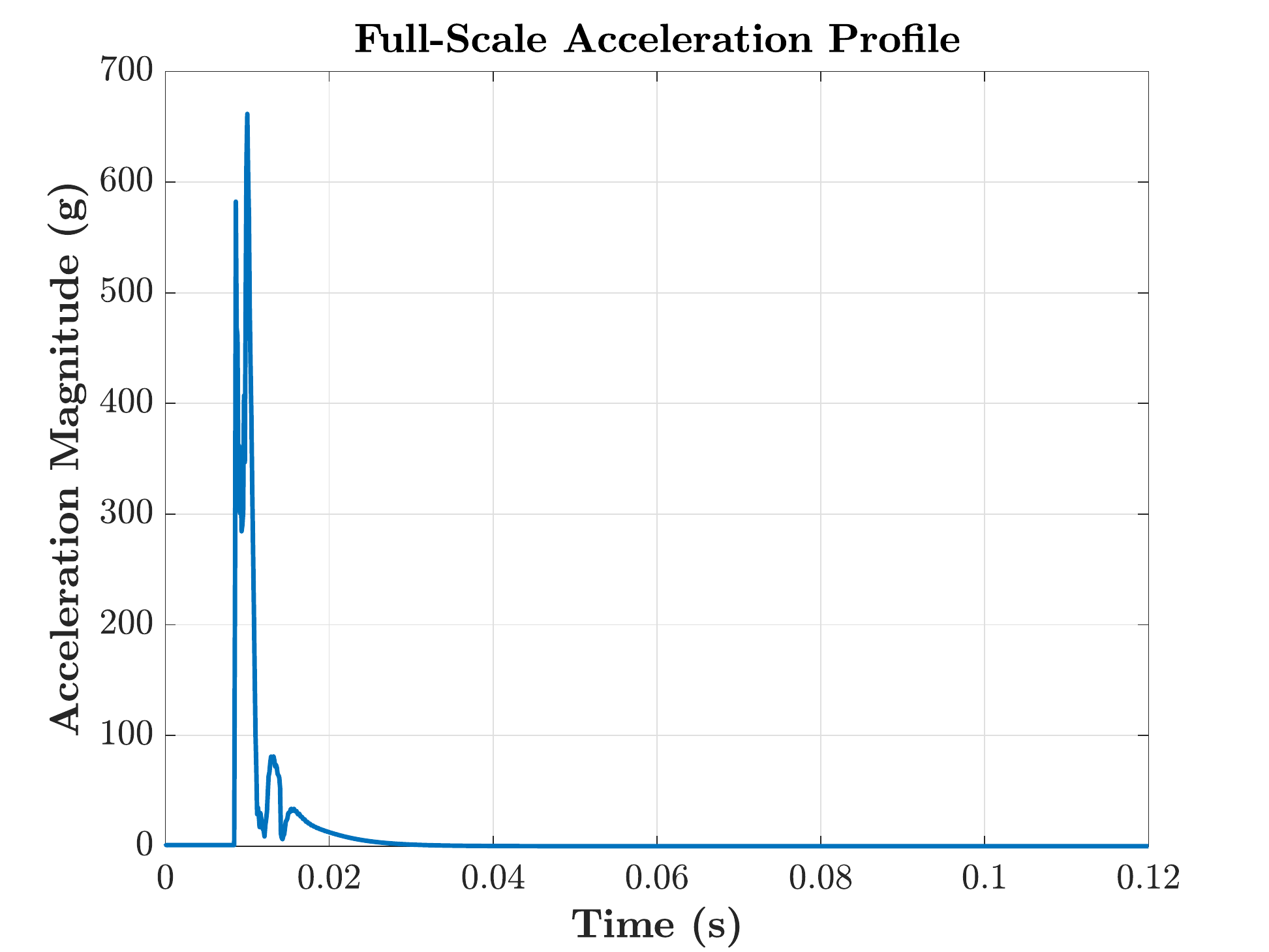}
    \caption{Full-scale acceleration profiles obtained from LS-Dyna simulations.}
    \label{fig: full_accel_profile}
\end{figure}

Figure \ref{fig: full_accel_profile} presents the acceleration profile from a simulated full-scale drop with an impact velocity of 10 $m/s$. 
A sizeable jump in the impact acceleration is noted with a damping behavior that is similar to the acceleration profiles of the small scale drops. 
It was the intention of the authors to pair these results with the experiment data from the full-scale drop test, but an issue with the batteries during these tests prevented the IMU data from being obtained. 

\section{Conclusions and Future Work}\label{sec: Conclusions}

The presented work provides an impact analysis of EagleCam, a payload for the Intuitive Machines IM-1 mission, upon landing on the lunar surface. 
The payload Concept of Operations is briefly discussed to provide context to the ensuing free-fall and impact analysis.
From these simulations, an anticipated impact velocity on the lunar surface is obtained. 
This value then serves as a driving force behind the structural design and testing to ensure survivability of the EagleCam payload upon impact. 

To verify survivability and validate impact simulations run in LS-Dyna, a series of small-scale drop tests were performed. 
The acceleration profiles of these tests are presented and compared with the simulation results.
It was concluded, that the simulated impacts were similar to those of the experimental drop tests with discrepancies attributed to disturbances and uncertainties in drop and impact. The simulation parameters will be further improved to match the experimental data by employing accelerometers with higher sampling frequency. 
Furthermore, a full-scale drop test was conducted to verify system survivability and results extrapolated to confirm simulations. 
A high-speed camera was used to verify impact velocity for the full-scale test.
During this experimentation the transmitted force to the internal structure made the batteries pop out of the power module, therefore it was not possible to obtain the data from IMU, but it will be redone in subsequent drops. 
Components were tested after the drop test and they operated nominally proving the internal force transmitted was mitigated and that the outer shell and the sand absorbed the majority of the impact energy.

In addition to obtaining supplementary IMU data, additional work will include improvements to both the simulation and experiments. 
Higher-fidelity modeling of the EagleCam payload and lunar surface will be considered through inclusion of the full internal assembly of the payload, composite layups varying Young's Modulus, finer meshing, and improved accuracy of energy dissipation to more closely match the impacting surface material. 
Sample lunar regolith simulant has been obtained and will be used as the material inside of the capture bay. 
As noted, the higher frequency resonance was missed by the IMU due to the low measurement frequency of the IMU.
For this reason, an IMU capable of sampling at higher frequencies will be used in subsequent experiments to capture the higher frequency phenomena of the impact.

\section{Acknowledgements}

The authors would like to gratefully acknowledge Dr. Ebenezer Gnanamanickam and Dr. Zheng Zhang for lending the high-speed camera for these experiments. 
The authors would also like to thank Jim Baker of Arrow Science and Technologies for manufacturing the outer shell of the drop module. 



\clearpage

\bibliographystyle{AAS_publication}   
\bibliography{references}   

\end{document}